\documentclass[review]{elsarticle}
\usepackage[margin=1in]{geometry}
\usepackage{graphicx}
\usepackage{amsmath}
\usepackage{amssymb}
\usepackage{subcaption}
\usepackage{multirow}
\usepackage{booktabs}
\usepackage[hidelinks]{hyperref}
\usepackage{float}

\title{Computer vision-based neural networks for radioisotope identification in urban environments}
\author[PNNLaddress]{Masen Bachleda\corref{corauthor}}
\author[PNNLaddress]{Alea Minar}
\author[PNNLaddress,UWashington]{Ayush Panigrahy}
\author[PNNLaddress]{Peter Lalor}
\date{June 2026}

\cortext[corauthor]{Corresponding author
\\\hspace*{13pt} Email address: masenbachleda@gmail.com
\\\hspace*{13pt} Telephone: (847) 346-6868}

\address[PNNLaddress]{Pacific Northwest National Laboratory, Richland, WA 99354 USA}
\address[UWashington]{University of Washington, Seattle, WA 98195 USA}

\begin{document}

\begin{abstract}
Algorithm development for radioisotope identification in mobile urban search scenarios face significant challenges from non-uniform backgrounds, momentary source encounters, and severe class imbalance between rare threat signatures and background measurements. We present a machine learning-based approach to this problem that converts list-mode gamma-ray data into two-dimensional waterfall spectrograms and applies computer vision architectures to the resulting images. Rather than treating waterfalls as conventional images, we employ a representation where consecutive time spectra can form input channels, similar to RGB channels in color images. This representation encodes both spectral and temporal information, enabling neural networks to more effectively learn patterns that distinguish source signatures from background fluctuations. We evaluate three architectures, a multilayer perceptron (MLP), convolutional neural network (CNN), and vision transformer (ViT), on the Radiological Anomaly Detection and Identification (RADAI) benchmark dataset. At a false positive rate of less than one false alarm per hour, our CNN outperforms the previous-best non-negative matrix factorization (NMF) method across all global metrics, achieving true detection, classification, and identification rates of 0.4334, 0.3965, and 0.2950 respectively, compared to 0.4151, 0.3611, and 0.2625 for NMF. At lower false positive rate constraints, the neural network approaches show comparable but ultimately lower performance than NMF, indicating opportunities for further research.

\begin{keyword}
Gamma Spectroscopy \sep Radioisotope Identification \sep Neural Networks \sep Waterfall Spectrogram
\end{keyword}

\end{abstract}

\maketitle

\section{Introduction}

Gamma-ray spectroscopy is a fundamental characterization method for identifying the isotopic composition of materials, with applications spanning nuclear physics, medical imaging, environmental monitoring, and nuclear security~\cite{knoll2010radiation}. By measuring the energy distribution of gamma photons emitted during radioactive decay, spectroscopic detectors produce characteristic spectra that can be matched to known isotope signatures. Locating and identifying radioactive materials that may be lost, stolen, or diverted for illicit purposes represents a critical national security capability, and mobile gamma-ray detection systems, typically using medium-resolution scintillators such as thallium-loaded sodium iodide (NaI(Tl)), are routinely deployed for urban radiological search~\cite{bilton2021neural, radai_dataset}. Urban environments present significant challenges for radiation detection algorithms~\cite{ghawaly2020chameleon, radai_dataset}. Naturally occurring radioactive material (NORM) in building materials, concrete, and soil creates location-dependent and temporally varying background that can obscure threat signatures. The combination of high background variability and transient source signatures makes urban radiological search a particularly demanding application for automated detection and identification systems.

Traditional approaches to gamma-ray spectroscopy rely on peak-based analysis, where characteristic photopeaks are identified and matched to known decay energies~\cite{knoll2010radiation}. While effective for high signal-to-noise measurements with extended integration times, these methods struggle in mobile search scenarios where observation times are short, backgrounds are dynamic, and multiple overlapping signatures may be present. More recently, machine learning methods have emerged as promising alternatives, with neural networks demonstrating the ability to learn optimal spectral representations for isotope classification tasks~\cite{kamuda2020comparison, daniel2020automatic, gomez2021isotope}. However, these methods typically operate on accumulated spectra integrated over fixed time periods, treating classification as a static pattern recognition problem and potentially discarding temporal information that could distinguish transient source encounters from background fluctuations.

Developing and rigorously evaluating detection algorithms requires large, labeled datasets that adequately cover the phase space of operational conditions~\cite{radai_dataset}. Acquiring real-world data is expensive, time-consuming, and fundamentally limited by safety constraints and environmental unpredictability. Furthermore, true ground truth is often unavailable in experimental measurements, complicating algorithm assessment and comparison. Machine learning methods are especially dependent on large training corpora, making synthetic data generation an attractive alternative~\cite{ghawaly2020chameleon, nicholson2020chameleon}. The Radiological Anomaly Detection and Identification (RADAI) dataset\footnote{\url{https://bdc.lbl.gov/wiki/public/radai-interactive-datasets/}}~\cite{radai_dataset} provides a standardized dataset for urban search algorithm development, featuring high-fidelity Monte Carlo simulations of mobile detection scenarios with detailed ground-truth labels. The dataset presents two key challenges for algorithm development: the temporal dynamics of mobile detection, where source encounters are brief events that must be detected within continuous measurements, and severe class imbalance, where background samples vastly outnumber source encounters.

In this work, we developed a methodology that takes the raw photon events and bins them into two-dimensional waterfall spectrograms, and then applies computer vision architectures to the subsequent images. Rather than treating the waterfall as a conventional image with spatial dimensions, we employ an unconventional representation where consecutive time spectra form the input channels, analogous to the RGB channels in color images. This design allows the neural networks to learn spectral patterns along the energy axis while preserving temporal information across channels. We evaluate three architectures, a multilayer perceptron (MLP), convolutional neural network (CNN), and Vision Transformer (ViT), on the RADAI benchmark, optimizing model performance using the True Identification Rate (TIDR) as our primary metric while constraining the False Positive Rate (FPR) to acceptable levels.

The remainder of this paper is organized as follows: Section~2 reviews related work in machine learning for gamma-ray spectroscopy and mobile radiation detection. Section~3 describes our methodology, including the waterfall preprocessing pipeline, neural network architectures, and evaluation framework. Section~4 presents experimental results on the RADAI benchmark, comparing our approach against baseline methods. Section~5 discusses the implications of our findings, and Section~6 provides concluding remarks.

\section{Related Work}

Machine learning methods have been extensively applied to gamma-ray spectroscopy for isotope identification~\cite{olmos1991automatic, keller1994gamma, yoshida2002neural, medhat2012artificial, aitkenhead2012plutonium, sheinfeld2017deep, daniel2020automatic}. Early work by Kamuda and Sullivan~\cite{kamuda2019automated} developed automated algorithms for identifying isotope mixtures in low-resolution spectra, while Kamuda et al.~\cite{kamuda2020comparison} provided a comprehensive comparison of machine learning approaches including neural networks, support vector machines, and decision trees for gamma-ray spectroscopy. Chen and Wei~\cite{chen2009nuclide} combined K-L transform with neural networks for nuclide identification, demonstrating the effectiveness of dimensionality reduction paired with machine learning approaches. Deep learning approaches have shown particular promise: Gomez-Fernandez et al.~\cite{gomez2021isotope} applied convolutional neural networks to isotope identification with explainability analysis, demonstrating that CNNs learn to focus on spectral regions corresponding to characteristic gamma-ray photopeaks rather than spurious features. Lalor et al.~\cite{lalor2026s2r, lalor2026unsuperviseddomainadaptationradioisotope} leverage domain adaptation techniques to efficiently utilize both simulated and experimental data for model training, achieving strong performance using a transformer-based neural network. These methods typically operate on accumulated spectra integrated over fixed time periods, treating the classification task as a static pattern recognition problem.

Mobile radiation detection introduces additional challenges due to the temporal dynamics of source encounters. As a detector moves through an environment, the observed spectrum varies continuously based on proximity to sources, changing backgrounds, and detector motion. Bilton et al.~\cite{bilton2021neural} specifically addressed mobile spectroscopic gamma-ray source detection using neural networks, demonstrating improved performance over traditional methods by incorporating temporal context. The Autoencoder Radiation Anomaly Detection (ARAD) model~\cite{ghawaly2022arad} employed autoencoders to learn background representations and detect anomalies in time-varying spectral data. Similarly, the gross-count k-sigma methods~\cite{archer2015ksigma} detect anomalies based on deviations from expected count rates. The RADAI dataset~\cite{radai_dataset} provides a standardized benchmark for evaluating radiation detection algorithms in urban search scenarios, building upon the earlier Chameleon Street dataset~\cite{ghawaly2020chameleon, nicholson2020chameleon}. Two baseline algorithms have been evaluated on this benchmark. The first one being the multiplexed censored energy window (mCEW)~\cite{lei2016mcew} algorithm, which optimizes energy windows for source detection and identification. The second one being the non-negative matrix factorization (NMF)~\cite{bilton2019nmf} algorithm, which will be discussed in the following paragraph. Among these baseline methods, NMF demonstrated the best overall performance on RADAI, while mCEW performed poorly potentially due to the high variability of urban backgrounds.

Non-negative matrix factorization (NMF), originally developed for applications in facial recognition and text mining~\cite{lee1999nmf}, has emerged as a leading approach for gamma-ray source detection and identification in mobile search scenarios. First applied to gamma-ray spectroscopy by Bilton et al.~\cite{bilton2019nmf}, NMF decomposes observed spectra into a linear combination of non-negative basis components representing learned spectral templates and activation weights. The non-negativity constraint is physically meaningful for gamma spectroscopy since photon counts cannot be negative, yielding interpretable ``parts-based'' representations where each basis component corresponds to recognizable spectral features such as the $^{40}$K 1460 keV line or uranium/thorium series contributions. Unlike principal component analysis (PCA), which produces orthogonal components with both positive and negative values lacking physical interpretation, NMF components can be directly understood as additive contributions from distinct radiation sources. NMF offers several key advantages: it utilizes full-spectrum information including the Compton scatter continuum rather than relying solely on isolated photopeaks, provides a compact low-dimensional representation reducing hundreds of spectral bins to just two or three basis components, and enables computationally efficient real-time inference once trained. For source identification, the background model is augmented with source templates, and a likelihood ratio test determines whether including the source template significantly improves the spectral fit, enabling simultaneous detection and identification within a unified statistical framework. However, NMF has notable limitations that motivate the exploration of alternative approaches: performance depends critically on having representative source templates in the library, the background model is static once trained and cannot adapt to changing environmental conditions, and the factorization is non-unique with solutions depending on initialization. These constraints suggest that template-based methods may be complemented by approaches that learn flexible, data-driven representations capable of generalizing beyond predefined templates. Despite these limitations, NMF has demonstrated state-of-the-art performance on radiation detection benchmarks and serves as the primary baseline for evaluating the neural network methods presented in this work.

\section{Methodology}

The Radiological Anomaly Detection and Identification (RADAI) dataset~\cite{radai_dataset} is a large-scale synthetic dataset developed through a collaboration between Oak Ridge National Laboratory, Lawrence Berkeley National Laboratory, Los Alamos National Laboratory, and Louisiana State University. The dataset was created to support the development and rigorous evaluation of radiation detection algorithms for urban search scenarios, with particular emphasis on enabling machine learning approaches that require large, labeled training corpora. RADAI consists of 300 training runs and 300 testing runs, each approximately one hour in duration, simulating a mobile 2"$\times$4"$\times$16" NaI(Tl) scintillator detector traversing urban environments.

The simulation employs high-fidelity Monte Carlo methods to model both the urban environment and detector response. Ten distinct city block types, including commercial districts, a bridge, park, and tunnel, can be combined via an adjacency-constrained ``Never-Ending Street'' approach to generate diverse route configurations. Background radiation includes terrestrial naturally occurring radioactive material (NORM) in building materials and soil, cosmic-ray contributions, and time-varying rain-induced backgrounds from $^{214}$Pb and $^{214}$Bi washout. Radioactive sources span 72 configurations across four categories: industrial isotopes (e.g., $^{137}$Cs, $^{60}$Co, $^{192}$Ir), medical isotopes (e.g., $^{99m}$Tc, $^{18}$F, $^{131}$I), nuclear materials (e.g., depleted and highly enriched uranium, fuel-grade and weapons-grade plutonium), and NORM sources (e.g., $^{40}$K, $^{226}$Ra, $^{232}$Th). Sources are placed at varying signal-to-noise ratios designed to achieve approximately 50\% detection probability at one false alarm per eight hours for baseline algorithms. For complete details on the simulation methodology, source configurations, and dataset design, we refer readers to the full RADAI dataset documentation~\cite{radai_dataset}.

\subsection{Problem Formulation}

We formulate isotope identification as a single-label, multi-class classification problem with 25 classes: one background class and 24 source types. The model outputs a probability distribution over all 25 classes via softmax activation. The RADAI competition leader board and evaluation scripts require us to submit our top three predictions. We do this by selecting the top three non-background predictions along with their predicted probability as a confidence score, enabling partial credit for near-miss classifications within isotope families. Since background predictions are not considered during scoring, we only submit source encounters where the top prediction was non-background.

A fundamental challenge in radiological search is the extreme class imbalance inherent to the detection problem. In urban monitoring scenarios, radioactive sources are rare events embedded within hours of background measurements. For a typical configuration (1-second time window with 6 stacked spectra spanning 6 seconds total), background represents approximately 98.5\% of samples, yielding an imbalance ratio exceeding 65:1 between background and all source classes combined. Standard cross-entropy loss performs poorly under such conditions since the majority class samples dominate the loss. To address this challenge, we employ Focal Loss~\cite{lin2017focal}, which dynamically down-weights well-classified examples and focuses training on hard, misclassified samples. Additionally, during hyperparameter optimization, we use evaluation metrics specifically designed for this kind of detection problem.

\subsection{Data Preprocessing Pipeline}

List mode data consists of individual photon detection events, each characterized by an energy value $E_i$ (in keV) and a time difference (in microseconds since the previous event). The first step in our preprocessing pipeline converts this event stream into one-dimensional gamma spectra. We partition the measurement period into $N_t$ consecutive intervals of duration $\Delta t$, defined by the time grid $t_{\text{edges}} = \{t_0, t_1, \ldots, t_{N_t}\}$. Within each interval, we bin all photon events by their energy values to produce a gamma spectrum $S_j \in \mathbb{R}^{N_E}$, where $N_E$ is the number of energy bins. The energy bins employ a nonlinear (square-root) binning scheme~\cite{bandstra2023explaining, ghawaly2022arad} to provide enhanced resolution at lower energies where characteristic gamma-ray photopeaks are most prominent. Following this approach, the $k$-th energy bin edge (for $k = 0, 1, \ldots, N_E$) is defined as:
\begin{equation}
E_k = \left( \sqrt{E_{\min}} + \frac{\sqrt{E_{\max}} - \sqrt{E_{\min}}}{N_E} \cdot k \right)^2
\end{equation}
For our implementation, we use $E_{\min} = 10$ keV and $E_{\max} = 3000$ keV. The raw histogram counts are normalized by the bin widths to produce a spectral density. This normalization yields units of counts/(keV$\cdot$s), making the spectral densities comparable across different bin sizes and time windows. By stacking $N_w$ consecutive spectral densities vertically, we obtain a two-dimensional waterfall plot, where each row corresponds to one time window and each column corresponds to one energy bin. This 2D representation enables our application of computer vision-based models, with the vertical axis representing temporal progression and the horizontal axis representing spectral energy channels.

Importantly, the waterfall dimensions are not fixed but are treated as hyperparameters optimized during model training. Specifically, we tune three parameters that determine the waterfall structure:
\begin{itemize}
    \item \textbf{Time window duration} $\Delta t \in \{0.5, 1, 2, 4\}$ seconds: The duration of each individual spectrum.
    \item \textbf{Number of stacked spectra} $N_w \in \{3, 4, 5, 6, 10, 12, 15, 20\}$: The number of consecutive spectra forming each waterfall. This determines the total temporal context available to the model, spanning $N_w \times \Delta t$ seconds.
    \item \textbf{Number of energy bins} $N_E \in \{32, 64, 128, 256\}$: The spectral resolution of each individual spectrum.
\end{itemize}

For each waterfall sample, the label is assigned a particular source if the time of closest approach to that source occurs during the waterfall sample. If no source is present in any of the $N_w$ windows, the waterfall is assigned the background label. This labeling strategy creates severe class imbalance, as background samples vastly outnumber source encounters in the dataset. In the unlikely scenario that two sources are present in the same waterfall, we label the waterfall according to whichever source arrives first. To enable efficient hyperparameter search across these configurations, we pre-compute waterfall files at 0.1-second temporal resolution with 1024 energy bins. During training, these files are dynamically merged and downsampled to the desired $(\Delta t, N_w, N_E)$ configuration, avoiding redundant computation of the base histograms across trials.

\begin{figure*}[t]
    \centering
    \begin{subfigure}[b]{0.48\textwidth}
        \centering
        \includegraphics[width=\textwidth]{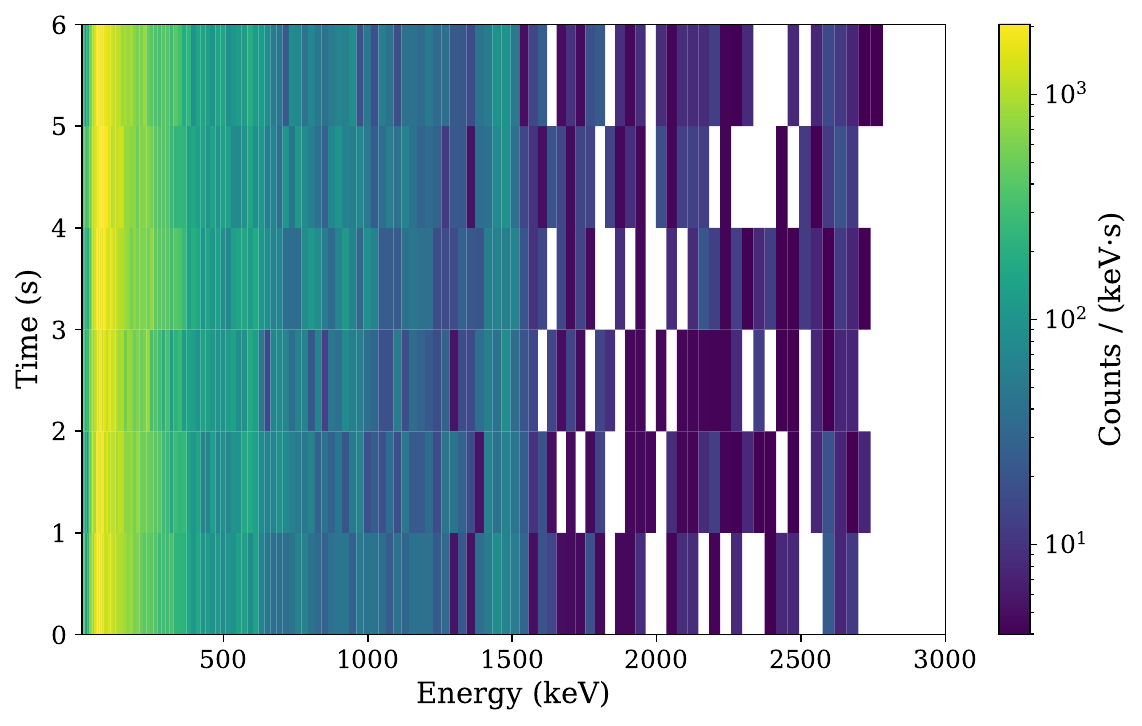}
        \caption{Background}
        \label{fig:waterfall_bkg}
    \end{subfigure}
    \hfill
    \begin{subfigure}[b]{0.48\textwidth}
        \centering
        \includegraphics[width=\textwidth]{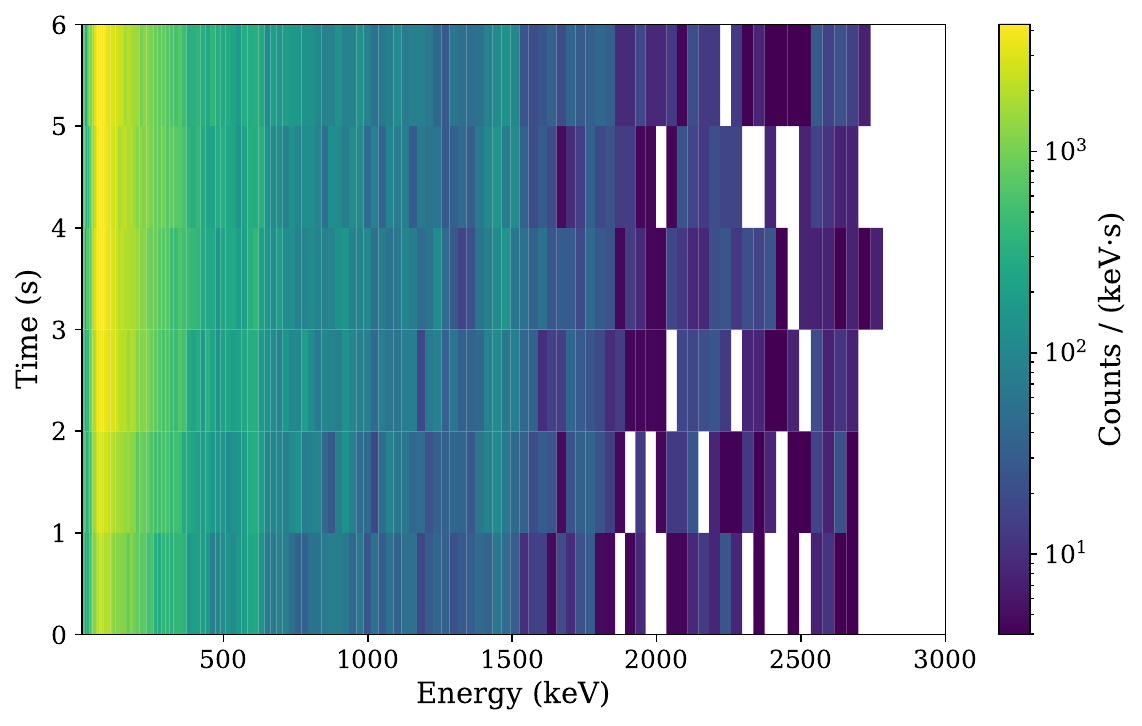}
        \caption{HEU source encounter}
        \label{fig:waterfall_heu}
    \end{subfigure}
    \caption{Example waterfall spectrograms showing (a) background radiation and (b) a highly enriched uranium (HEU) source encounter from the RADAI dataset. Each waterfall spans $N_w = 6$ time windows of $\Delta t = 1.0$ second duration with $N_E = 128$ energy bins. The vertical axis represents time progression, the horizontal axis shows energy (keV), and color intensity indicates normalized count rate.}
    \label{fig:waterfall_examples}
\end{figure*}

Rather than partitioning each run into non-overlapping waterfall samples, we slide an $N_w$-spectrum window through each run with a stride of one spectrum. This generates heavily overlapping waterfall samples, substantially increasing the effective training set size. For example, a 3600-second run with $\Delta t = 1$ second and $N_w = 6$ yields approximately 3595 overlapping windows instead of 600 non-overlapping samples. This strategy captures source encounters at different temporal positions within the waterfall, providing diverse training examples where sources may appear in early, middle, or late time windows.

The same sliding window approach is applied during inference on test data. The model generates predictions at every possible $N_w$-spectrum position, producing dense temporal sampling with predictions every $\Delta t$ seconds. This generates multiple overlapping predictions for each source encounter, so we employ post-processing to consolidate redundant alarms. We considered instead partitioning the test data into non-overlapping waterfalls to simplify the post-processing step. However, we found reduced sensitivity, especially when a source existed near the boundary of two waterfalls, and we thus favor a sliding-window and post-processing inference solution.

\subsection{Evaluation Metrics}

We adopt the evaluation framework established by the RADAI challenge~\cite{radai_dataset}, which defines metrics appropriate for time-series anomaly detection. The primary metrics are:

\begin{description}
\item[True Detection Rate (TDR)] measures the fraction of source encounters that are detected, regardless of whether the isotope is correctly classified:
\begin{equation}
\text{TDR} = \frac{\text{Number of detected source encounters}}{\text{Total number of source encounters}}
\end{equation}
A source encounter is considered ``detected'' if the model predicts the presence of a source (non-background prediction) within a specified time window around the true source presence, defined as $\pm$80 seconds from the time of closest approach.

\item[True Classification Rate (TCR)] measures the fraction of detected sources that are correctly classified:
\begin{equation}
\text{TCR} = \frac{\text{Number of correctly classified detections}}{\text{Number of detected source encounters}}
\end{equation}
This metric evaluates classification accuracy conditional on successful detection.

\item[True Identification Rate (TIDR)] is the primary optimization target, combining detection and classification into a single metric:
\begin{equation}
\text{TIDR} = \frac{\text{Number of correctly identified source encounters}}{\text{Total number of source encounters}} = \text{TDR} \times \text{TCR}
\end{equation}
This stringent metric captures the end-to-end performance relevant to operational scenarios where both detection and identification are required.

\item[False Positive Rate (FPR)] measures the rate of false alarms in background regions:
\begin{equation}
\text{FPR} = \frac{\text{Number of false alarms}}{\text{Total background observation time}}
\end{equation}
In the RADAI framework, FPR is normalized by total background observation time rather than by sample count, yielding a rate with units of alarms per unit time. This formulation is operationally meaningful: an FPR of 1 corresponds to one false alarm per hour of background observation.

\end{description}

We use TIDR as our evaluation criterion during hyperparameter optimization rather than TDR and TCR because TIDR captures the complete identification task. A system with high TDR but low TCR would generate many detections but frequently misidentify isotopes, which may be operationally worse than missing some sources entirely. Conversely, optimizing TCR alone could bias the model toward conservative predictions, harming detection sensitivity.

\subsection{Submission Generation Pipeline}

Converting model predictions into a final submission requires a two-stage post-processing pipeline that addresses the redundancy introduced by sliding window evaluation and enforces the FPR constraint.

\begin{description}

\item[Stage 1: Temporal Clustering.] The model generates predictions at every possible temporal position with stride $\Delta t$. For a single source encounter, this produces multiple overlapping detections (potentially dozens of alarms within a few seconds). To consolidate these redundant predictions, we apply time-based clustering with a 20-second threshold. Specifically, consecutive non-background predictions within 20 seconds of each other are grouped into clusters, and only the prediction with the highest confidence score is retained from each cluster. This threshold was chosen based on the RADAI dataset design, where source encounters are separated by at least 30 seconds, ensuring that the 20-second clustering window captures only redundant detections of the same source rather than merging distinct encounters.

\item[Stage 2: FPR-Constrained Threshold Selection.] After clustering, we perform a threshold search to identify the confidence threshold that maximizes TIDR while satisfying the FPR constraint. Predictions below threshold are suppressed (treated as background). We select the lowest possible threshold in order to satisfy the FPR requirement for each submission category (FPR $<$ 0.125, FPR $<$ 0.25, FPR $<$ 1). Importantly, predictions are computed only once; the threshold search operates on cached softmax outputs, making this sweep computationally inexpensive.

\end{description}

The final submission is a CSV file containing one row per alarm, with columns specifying the run identifier, alarm time window, and the top three non-background class predictions with their confidence scores.

\subsection{Neural Network Architectures}

We evaluate three neural network architectures representing different inductive biases for processing waterfall spectrograms: a multilayer perceptron (MLP), a one-dimensional convolutional neural network (CNN), and a vision transformer (ViT). The input dimensions $(N_w, N_E)$ are jointly optimized with model hyperparameters, requiring all architectures to handle variable input sizes ranging from $(3, 32)$ to $(20, 256)$. All architectures share a common preprocessing layer that applies a square-root variance-stabilizing transform followed by per-sample Z-score normalization, which stabilizes variance for Poisson-distributed count data and ensures consistent input scales across different detector conditions. Below, we motivate each architecture by the assumptions it encodes about spectral data.

\begin{description}

\item[Multilayer Perceptrons] flatten the waterfall into a single feature vector and process it through fully-connected layers with Layer Normalization~\cite{ba2016layer}, ReLU activations, and dropout. This architecture imposes no spatial or temporal structure, testing whether learned feature relationships alone suffice for isotope identification. However, the absence of inductive biases means MLPs must encounter examples of all realistic spectral variations during training to generalize effectively. In gamma spectroscopy, this manifests as potential difficulty generalizing across unseen detector calibrations or source configurations.

\item[Convolutional Neural Networks] incorporate one-dimensional convolutional layers that slide learnable filters across local segments of the spectrum, exploiting the assumption that information from adjacent energy channels is correlated. This inductive bias is well-suited to gamma spectroscopy, where characteristic photopeaks span multiple adjacent energy bins due to detector resolution. Our architecture transposes the input waterfall so that 1D convolutions operate along the energy axis while treating the $N_w$ time windows as input channels, resembling color channels in image CNNs. This design enables the convolutional filters to learn spectral patterns (e.g., photopeak shapes) that are consistent across time. Each convolutional block applies Batch Normalization~\cite{ioffe2015batch}, ReLU activation, and max pooling, followed by fully-connected classification layers with dropout.

\item[Vision Transformers] adapt the transformer's self-attention mechanism to process waterfall plots~\cite{dosovitskiy2021vit}. Unlike the CNN's local receptive fields, self-attention enables each position to attend to all other positions, potentially capturing long-range correlations between distant spectral features, for example, identifying isotopes by jointly considering widely separated photopeaks. Following the approach of Lalor et al.~\cite{lalor2026s2r, lalor2026unsuperviseddomainadaptationradioisotope} for gamma spectroscopy, and mirroring our CNN's design philosophy, we apply a 1D patching strategy along the energy axis where each patch spans all $N_w$ time windows. Patches are linearly projected to an embedding space with learnable positional encodings, then processed by transformer blocks using pre-norm multi-head self-attention and feedforward layers with residual connections. Global average pooling aggregates the patch embeddings into a single representation for classification.

\end{description}

\subsection{Experimental Setup}

All models are trained using the AdamW optimizer with Focal Loss~\cite{lin2017focal} to address class imbalance, early stopping (patience of 7 epochs), and learning rate reduction on plateau. Models were trained on the 300 RADAI training runs using a 70/15/15 train/validation/test split (210/45/45 runs). Hyperparameter optimization was performed using Optuna~\cite{akiba2019optuna} with 300 trials and a Tree-structured Parzen Estimator sampler, maximizing validation TIDR at the target FPR constraint. Waterfall representation parameters $(N_w, N_E, \Delta t)$ are optimized jointly with model hyperparameters; complete hyperparameter search spaces are provided in~\ref{sec:hyperparams}. Final evaluation was performed by generating a submission file using the best model and uploading to the RADAI competition leaderboard.

\section{Results}

We evaluated the performance of the multilayer perceptron (MLP), convolutional neural network (CNN), and vision transformer (ViT) against the baseline non-negative matrix factorization (NMF) method on the RADAI dataset. Table~\ref{tab:global_results} presents the global performance metrics across three false positive rate (FPR) constraint levels. At the false positive rate constraint of less than one false alarm per hour (FPR $<$ 1), the CNN achieved the highest performance across all three global metrics, with a true detection rate (TDR) of 0.4334, true classification rate (TCR) of 0.3965, and true identification rate (TIDR) of 0.2950. This represents a meaningful improvement over the baseline NMF method, which achieved a TDR of 0.4151, TCR of 0.3611, and TIDR of 0.2625.

\begin{table}[H]
\centering
\caption{Global performance metrics for different neural network architectures across three FPR constraint levels. Bold values indicate the best performance for each metric.}
\label{tab:global_results}

\begin{subtable}{\textwidth}
\centering
\caption{Global results at FPR $<$ 1.0}
\label{tab:global_fpr_1}
\begin{tabular}{lccc}
\toprule
\textbf{Method} & \textbf{TDR} & \textbf{TCR} & \textbf{TIDR} \\
\midrule
CNN & \textbf{0.4334} & \textbf{0.3965} & \textbf{0.2950} \\
MLP & 0.2884 & 0.2603 & 0.2118 \\
ViT & 0.3687 & 0.3300 & 0.2300 \\
Baseline NMF & 0.4151 & 0.3611 & 0.2625 \\
\bottomrule
\end{tabular}
\end{subtable}

\vspace{0.5em}

\begin{subtable}{\textwidth}
\centering
\caption{Global results at FPR $<$ 0.25}
\label{tab:global_fpr_025}
\begin{tabular}{lccc}
\toprule
\textbf{Method} & \textbf{TDR} & \textbf{TCR} & \textbf{TIDR} \\
\midrule
CNN & 0.2096 & 0.1913 & 0.1738 \\
MLP & 0.1818 & 0.1654 & 0.1438 \\
ViT & 0.1544 & 0.1365 & 0.1289 \\
Baseline NMF & \textbf{0.3348} & \textbf{0.2968} & \textbf{0.2234} \\
\bottomrule
\end{tabular}
\end{subtable}

\vspace{0.5em}

\begin{subtable}{\textwidth}
\centering
\caption{Global results at FPR $<$ 0.125}
\label{tab:global_fpr_0125}
\begin{tabular}{lccc}
\toprule
\textbf{Method} & \textbf{TDR} & \textbf{TCR} & \textbf{TIDR} \\
\midrule
CNN & 0.1617 & 0.1479 & 0.1391 \\
MLP & 0.1296 & 0.1165 & 0.1062 \\
ViT & 0.1292 & 0.1139 & 0.1106 \\
Baseline NMF & \textbf{0.2979} & \textbf{0.2643} & \textbf{0.2026} \\
\bottomrule
\end{tabular}
\end{subtable}

\end{table}

Among the neural network architectures, a clear performance hierarchy emerges at the FPR $<$ 1 constraint: the CNN achieves the highest TIDR (0.2950), followed by the ViT (0.2300), and then the MLP (0.2118). Part of the CNN's advantage may stem from its detection capability, achieving a TDR nearly 50\% higher than the MLP (0.4334 vs.\ 0.2884). The ViT falls between these two, with intermediate performance across all metrics. Notably, all three architectures demonstrate different trade-offs between detection sensitivity and classification accuracy: the CNN maintains a high TCR relative to its TDR (0.3965/0.4334 = 91\%), while the MLP shows a similar ratio (0.2603/0.2884 = 90\%) at lower absolute values.

However, the relative performance of these methods diverges at lower false positive rate constraints. At the constraint of less than one false alarm per four hours (FPR $<$ 0.25), the baseline NMF achieves a TIDR of 0.2234, substantially outperforming the CNN (0.1738), MLP (0.1438), and ViT (0.1289). This gap widens further at the most stringent constraint of less than one false alarm per eight hours (FPR $<$ 0.125), where NMF achieves a TIDR of 0.2026 compared to 0.1391 for the CNN. This pattern suggests that while neural networks can achieve higher performance when moderate false alarm rates are acceptable, the template-based NMF approach maintains more robust performance under tighter constraints where false alarms are minimal.

The global metrics aggregate performance across four distinct source categories, Industrial, Medical, Naturally Occurring Radioactive Material (NORM), and Nuclear Material, each presenting different detection challenges. Table~\ref{tab:category_results} provides category-specific metrics with global FPR $<$ 1, revealing substantial variation in how each method performs across source types. Category-specific results at lower FPR constraints can be found in~\ref{sec:category_results_low_fpr}.

\begin{table}[H]
\centering
\caption{Category-specific performance metrics with global FPR $<$ 1.0 for different neural network architectures. Bold values indicate the best performance for each metric.}
\label{tab:category_results}

\begin{subtable}{\textwidth}
\centering
\caption{Industrial and Medical results with global FPR $<$ 1.0}
\label{tab:category_ind_med}
\begin{tabular}{lcccccccc}
\toprule
& \multicolumn{4}{c}{\textbf{Industrial}} & \multicolumn{4}{c}{\textbf{Medical}} \\
\cmidrule(lr){2-5} \cmidrule(lr){6-9}
\textbf{Method} & \textbf{TDR} & \textbf{TCR} & \textbf{TIDR} & \textbf{FPR} & \textbf{TDR} & \textbf{TCR} & \textbf{TIDR} & \textbf{FPR} \\
\midrule
CNN & 0.4661 & 0.3939 & 0.3939 & 0.0524 & \textbf{0.3758} & 0.2527 & 0.2376 & \textbf{0.1187} \\
MLP & 0.3611 & 0.3042 & 0.3042 & \textbf{0.0209} & 0.3132 & 0.2225 & 0.2181 & 0.2653 \\
ViT & 0.4201 & 0.3435 & 0.3435 & 0.1815 & 0.2678 & 0.1339 & 0.1253 & 0.1466 \\
Baseline NMF & \textbf{0.6280} & \textbf{0.5777} & \textbf{0.5755} & 0.3072 & 0.3693 & \textbf{0.2549} & \textbf{0.2549} & 0.2199 \\
\bottomrule
\end{tabular}
\end{subtable}

\vspace{0.5em}

\begin{subtable}{\textwidth}
\centering
\caption{NORM and Nuclear Material results with global FPR $<$ 1.0}
\label{tab:category_norm_nuc}
\begin{tabular}{lcccccccc}
\toprule
& \multicolumn{4}{c}{\textbf{NORM}} & \multicolumn{4}{c}{\textbf{Nuclear Material}} \\
\cmidrule(lr){2-5} \cmidrule(lr){6-9}
\textbf{Method} & \textbf{TDR} & \textbf{TCR} & \textbf{TIDR} & \textbf{FPR} & \textbf{TDR} & \textbf{TCR} & \textbf{TIDR} & \textbf{FPR} \\
\midrule
CNN & \textbf{0.0965} & \textbf{0.0921} & \textbf{0.0921} & 0.4294 & \textbf{0.4890} & \textbf{0.4827} & \textbf{0.3124} & 0.3840 \\
MLP & 0.0219 & 0.0219 & 0.0219 & 0.3037 & 0.2986 & 0.2929 & 0.2106 & \textbf{0.3316} \\
ViT & 0.0175 & 0.0044 & 0.0044 & \textbf{0.0524} & 0.4337 & 0.4299 & 0.2602 & 0.6144 \\
Baseline NMF & 0.0044 & 0.0044 & 0.0044 & 0.0873 & 0.4261 & 0.3809 & 0.2118 & 0.3805 \\
\bottomrule
\end{tabular}
\end{subtable}

\end{table}

For Industrial sources, which include isotopes with distinctive spectral signatures such as $^{137}$Cs (662 keV photopeak) and $^{60}$Co (1173/1332 keV photopeaks), the baseline NMF achieves the highest identification rate (TIDR = 0.5755), outperforming all neural network approaches. The CNN achieves a TIDR of 0.3939, while maintaining a much lower false positive rate (0.0524 vs.\ 0.3072 for NMF). For Medical sources, which include short-lived isotopes such as $^{99m}$Tc and $^{18}$F, performance is more comparable: NMF achieves a TIDR of 0.2549 compared to 0.2376 for the CNN, with the CNN showing a lower false positive rate (0.1187 vs.\ 0.2199).

The most striking differences between methods emerge for NORM and Nuclear Material categories. NORM sources ($^{40}$K, $^{226}$Ra, and $^{232}$Th) present the greatest detection challenge because their spectra closely resemble terrestrial background radiation. Here, the CNN achieves a detection rate of 0.0965 compared to just 0.0044 for NMF, a more than 20-fold improvement, though detection remains difficult for all methods. This improvement comes with a trade-off: the CNN's NORM false positive rate (0.4294) is higher than NMF's (0.0873), reflecting a more aggressive detection strategy. For Nuclear Material sources, the CNN achieves the highest identification rate (TIDR = 0.3124) compared to 0.2118 for NMF, representing a 47\% relative improvement. The ViT also performs competitively on Nuclear Materials (TIDR = 0.2602), suggesting that its attention mechanism may be well-suited for capturing the complex multi-peak signatures characteristic of nuclear material.

\section{Discussion}

The per-category results illuminate how different source types interact with each detection approach. The CNN's improvement over NMF for NORM sources (TDR of 0.0965 versus 0.0044) represents the most striking finding. NORM isotopes ($^{40}$K, $^{226}$Ra, and $^{232}$Th) produce spectra that closely resemble terrestrial background radiation, making them exceptionally difficult to detect via template matching. The CNN's learned convolutional filters appear capable of identifying subtle deviations from expected background patterns that fixed NMF templates cannot capture. Conversely, NMF's superior performance on Industrial sources (TDR of 0.6280 versus 0.4661 for CNN) suggests that template-based approaches excel when sources exhibit distinctive, well-separated photopeaks. Industrial isotopes such as $^{137}$Cs (662 keV) and $^{60}$Co (1173/1332 keV) produce unambiguous spectral signatures that are easily matched by reference templates; the neural networks may be learning unnecessary complexity for these straightforward classification tasks. For Nuclear Materials, the CNN's improvement (TIDR of 0.3124 versus 0.2118 for NMF) indicates that learned features capture information beyond simple peak identification, possibly including Compton scatter continuum shapes or correlations among the multiple gamma-ray lines characteristic of nuclear decay chains.

These findings suggest that the optimal detection strategy depends on the operational context. Template-based methods like NMF remain highly effective for sources with distinctive spectral signatures, particularly industrial isotopes. However, neural network approaches, especially CNNs, demonstrate clear advantages for challenging detection scenarios where subtle learned features outperform hand-crafted templates. The CNN's strong performance on NORM and Nuclear Material categories motivates further investigation into hybrid approaches or ensemble methods that could combine the complementary strengths of template-based and learned representations.
    
\section{Conclusion}

This work investigated machine learning methods for radioisotope identification in urban search scenarios. We converted list-mode gamma-ray data from the RADAI benchmark dataset into waterfall spectrograms and trained three computer vision-based architectures on the corresponding images. At the FPR constraint of less than one false alarm per hour, our CNN achieved a global TIDR of 0.2950, and outperformed the baseline NMF method in all three global metrics. The CNN showed particular strength for challenging detection scenarios: NORM sources saw a significant improvement in detection rate, and Nuclear Material sources achieved a TIDR of 0.3124 compared to 0.2118 for NMF. These results demonstrate that learned convolutional features can capture subtle spectral patterns that template-based methods miss. However, at lower FPR constraints, the baseline NMF currently outperforms all neural network approaches, indicating room for improvement of the neural network models. Several directions warrant future investigation. A two-stage approach that separates detection from classification could improve performance by allowing specialized models for each task. Data augmentation strategies, such as synthetic source injection, may help neural networks generalize better across varying detector conditions. Ensemble methods combining multiple architectures, or hybrid approaches that integrate template matching with learned representations, could leverage the complementary strengths observed in our analysis. Our results show that machine learning methods are competitive with and can outperform traditional template-based approaches for radiation detection. As benchmark datasets like RADAI continue to drive algorithm development, we expect neural network methods to play an increasing role in radiation detection systems.

\section{Declaration of Generative AI and AI-assisted technologies in the writing process}

During the preparation of this work the author(s) used generative AI tools (Claude Sonnet 3.5/Claude Opus 4.5, Anthropic) in order to assist with language editing and formatting. After using this tool/service, the author(s) reviewed and edited the content as needed and take(s) full responsibility for the content of the published article.

\section{Declaration of competing interest}
The authors declare that they have no known competing financial interests or personal relationships that could have appeared to influence the work reported in this paper.

\section{Acknowledgements}

This research was supported by the Laboratory Directed Research and Development Program at Pacific Northwest National Laboratory, a multiprogram national laboratory operated by Battelle for the U.S. Department of Energy under contract DE-AC05-76RLO1830. Masen Bachleda is grateful for the support of the Linus Pauling Distinguished Postdoctoral Fellowship. The authors would like to acknowledge Maria Glenski for her useful suggestions and feedback.

\newpage

\setlength{\bibsep}{0pt}
\bibliography{References.bib}

\newpage

\appendix

\section{Hyperparameter Search Spaces}
\label{sec:hyperparams}

This appendix provides the complete hyperparameter search spaces used during Optuna optimization for each architecture. Table~\ref{tab:hyperparams} details the search spaces and best-performing hyperparameter values obtained via Bayesian optimization. Waterfall representation parameters (time window duration, number of stacked spectra, and number of energy bins) are optimized jointly with model hyperparameters. Curly braces indicate categorical choices; square brackets indicate continuous ranges ($\dagger$ denotes log-scale sampling).

\begin{table}[h]
\centering
\caption{Hyperparameter optimization summary for MLP, CNN, and Vision Transformer architectures. The table details the search spaces and best-performing hyperparameter values obtained via Bayesian optimization.}
\label{tab:hyperparams}
\small
\begin{tabular}{llcc}
\toprule
\textbf{Architecture} & \textbf{Parameter} & \textbf{Search Space} & \textbf{Best run} \\
\midrule
\multirow{12}{*}{MLP} 
    & Time window $\Delta t$ (s) & \{0.5, 1, 2, 4\} & 1 \\
    & Number of windows $N_w$ & \{3, 4, 5, 6, 10, 12, 15, 20\} & 5 \\
    & Energy bins $N_E$ & \{32, 64, 128, 256\} & 64 \\
    & Learning rate$^\dagger$ & $[10^{-5}, 5 \times 10^{-5}]$ & 1.22e-5 \\
    & Batch size & \{16, 32, 64, 128, 256\} & 128 \\
    & Weight decay$^\dagger$ & $[10^{-7}, 5 \times 10^{-5}]$ & 1.05e-7 \\
    & Focal Loss $\alpha$ & $[0.1, 0.5]$ & 0.440 \\
    & Focal Loss $\gamma$ & $[1.0, 3.0]$ & 1.29 \\
    & Number of hidden layers & \{1, 2, 3, 4, 5\} & 3 \\
    & Base layer size & \{128, 256, 512, 1024, 2048\} & 2048 \\
    & Dropout rate & $[0.0, 0.5]$ & 0.137 \\
\midrule
\multirow{15}{*}{CNN} 
    & Time window $\Delta t$ (s) & \{0.5, 1, 2, 4\} & 0.5 \\
    & Number of windows $N_w$ & \{3, 4, 5, 6, 10, 12, 15, 20\} & 10 \\
    & Energy bins $N_E$ & \{32, 64, 128, 256\} & 128 \\
    & Learning rate$^\dagger$ & $[10^{-5}, 5 \times 10^{-5}]$ & 3.78e-5 \\
    & Batch size & \{16, 32, 64, 128, 256\} & 32 \\
    & Weight decay$^\dagger$ & $[10^{-7}, 5 \times 10^{-5}]$ & 3.21e-6 \\
    & Focal Loss $\alpha$ & $[0.1, 0.5]$ & 0.111 \\
    & Focal Loss $\gamma$ & $[1.0, 3.0]$ & 1.25 \\
    & Number of conv.\ layers & \{1, 2, 3, 4\} & 3 \\
    & Base filter count & \{16, 32\} & 32 \\
    & Kernel size & \{5, 7\} & 7 \\
    & Number of dense layers & \{1, 2, 3\} & 1 \\
    & Base dense units & \{512, 1024\} & 1024 \\
    & Dropout rate & $[0.02, 0.15]$ & 0.148 \\
\midrule
\multirow{14}{*}{ViT} 
    & Time window $\Delta t$ (s) & \{0.5, 1, 2, 4\} & 2 \\
    & Number of windows $N_w$ & \{3, 4, 5, 6, 10, 12, 15, 20\} & 3 \\
    & Energy bins $N_E$ & \{32, 64, 128, 256\} & 256 \\
    & Learning rate$^\dagger$ & $[10^{-5}, 5 \times 10^{-5}]$ & 4.29e-5 \\
    & Batch size & \{16, 32, 64, 128, 256\} & 16 \\
    & Weight decay$^\dagger$ & $[10^{-7}, 5 \times 10^{-5}]$ & 8.44e-7 \\
    & Focal Loss $\alpha$ & $[0.1, 0.5]$ & 0.489 \\
    & Focal Loss $\gamma$ & $[1.0, 3.0]$ & 1.17 \\
    & Patch size $p_E$ & \{16, 32, 64, 128, 256\} & 32 \\
    & Projection dimension $d$ & \{64, 128, 256, 384, 512\} & 64 \\
    & Transformer layers & \{1, 2, 3, 4, 5, 6\} & 6 \\
    & Attention heads $H$ & \{1, 2, 4, 8, 16\} & 16 \\
    & FF multiplier $m$ & \{2, 3, 4\} & 2 \\
    & Dropout rate & $[0.0, 0.5]$ & 0.047 \\
\bottomrule
\end{tabular}
\end{table}

\noindent\textbf{ViT Constraints.} $d \mod H = 0$ (projection dimension must be divisible by the number of attention heads); $N_E / p_E \geq 4$ (minimum 4 patches required). The feedforward dimension is computed as $d_{\text{ff}} = m \cdot d$.

\section{Category Results at Lower FPR Constraints}
\label{sec:category_results_low_fpr}

This appendix provides category-specific performance metric results with global FPR $<$ 0.25 and global FPR $<$ 0.125 constraint levels, complementing the results presented in Tables~\ref{tab:category_ind_med} and~\ref{tab:category_norm_nuc}.

\begin{table}[h]
\centering
\caption{Category-specific performance metrics at lower FPR constraints. Bold values indicate the best performance for each metric.}
\label{tab:category_results_low_fpr_table}

\begin{subtable}{\textwidth}
\centering
\caption{Industrial and Medical results with global FPR $<$ 0.25}
\label{tab:category_ind_med_025}
\begin{tabular}{lcccccccc}
\toprule
& \multicolumn{4}{c}{\textbf{Industrial}} & \multicolumn{4}{c}{\textbf{Medical}} \\
\cmidrule(lr){2-5} \cmidrule(lr){6-9}
\textbf{Method} & \textbf{TDR} & \textbf{TCR} & \textbf{TIDR} & \textbf{FPR} & \textbf{TDR} & \textbf{TCR} & \textbf{TIDR} & \textbf{FPR} \\
\midrule
CNN & 0.3348 & 0.3107 & 0.3107 & \textbf{0.0035} & 0.2505 & 0.1728 & 0.1706 & \textbf{0.0175} \\
MLP & 0.2604 & 0.2341 & 0.2341 & \textbf{0.0035} & 0.2268 & 0.1598 & 0.1555 & 0.0733 \\
ViT & 0.3020 & 0.2867 & 0.2867 & 0.0803 & 0.1836 & 0.0972 & 0.0929 & 0.0524 \\
Baseline NMF & \textbf{0.5492} & \textbf{0.5077} & \textbf{0.5055} & 0.1152 & \textbf{0.3067} & \textbf{0.2181} & \textbf{0.2181} & 0.0419 \\
\bottomrule
\end{tabular}
\end{subtable}

\vspace{0.5em}

\begin{subtable}{\textwidth}
\centering
\caption{NORM and Nuclear Material results with global FPR $<$ 0.25}
\label{tab:category_norm_nuc_025}
\begin{tabular}{lcccccccc}
\toprule
& \multicolumn{4}{c}{\textbf{NORM}} & \multicolumn{4}{c}{\textbf{Nuclear Material}} \\
\cmidrule(lr){2-5} \cmidrule(lr){6-9}
\textbf{Method} & \textbf{TDR} & \textbf{TCR} & \textbf{TIDR} & \textbf{FPR} & \textbf{TDR} & \textbf{TCR} & \textbf{TIDR} & \textbf{FPR} \\
\midrule
CNN & \textbf{0.0526} & \textbf{0.0482} & \textbf{0.0482} & 0.1571 & 0.1842 & 0.1829 & 0.1534 & 0.0628 \\
MLP & 0.0088 & 0.0088 & 0.0088 & 0.0628 & 0.1710 & 0.1697 & 0.1339 & 0.1082 \\
ViT & 0.0044 & 0.0000 & 0.0000 & \textbf{0.0175} & 0.1251 & 0.1245 & 0.1125 & 0.0977 \\
Baseline NMF & 0.0044 & 0.0044 & 0.0044 & 0.0279 & \textbf{0.3287} & \textbf{0.3011} & \textbf{0.1754} & \textbf{0.0628} \\
\bottomrule
\end{tabular}
\end{subtable}

\vspace{0.5em}

\begin{subtable}{\textwidth}
\centering
\caption{Industrial and Medical results with global FPR $<$ 0.125}
\label{tab:category_ind_med_0125}
\begin{tabular}{lcccccccc}
\toprule
& \multicolumn{4}{c}{\textbf{Industrial}} & \multicolumn{4}{c}{\textbf{Medical}} \\
\cmidrule(lr){2-5} \cmidrule(lr){6-9}
\textbf{Method} & \textbf{TDR} & \textbf{TCR} & \textbf{TIDR} & \textbf{FPR} & \textbf{TDR} & \textbf{TCR} & \textbf{TIDR} & \textbf{FPR} \\
\midrule
CNN & 0.2867 & 0.2691 & 0.2691 & \textbf{0.0000} & \textbf{0.2009} & 0.1361 & 0.1339 & 0.0070 \\
MLP & 0.2166 & 0.1947 & 0.1947 & \textbf{0.0000} & 0.1728 & 0.1210 & 0.1166 & 0.0384 \\
ViT & 0.2582 & 0.2473 & 0.2473 & 0.0384 & 0.1598 & 0.0799 & 0.0778 & 0.0105 \\
Baseline NMF & \textbf{0.5098} & \textbf{0.4705} & \textbf{0.4705} & 0.0698 & 0.2700 & \textbf{0.1965} & \textbf{0.1965} & \textbf{0.0035} \\
\bottomrule
\end{tabular}
\end{subtable}

\vspace{0.5em}

\begin{subtable}{\textwidth}
\centering
\caption{NORM and Nuclear Material results with global FPR $<$ 0.125}
\label{tab:category_norm_nuc_0125}
\begin{tabular}{lcccccccc}
\toprule
& \multicolumn{4}{c}{\textbf{NORM}} & \multicolumn{4}{c}{\textbf{Nuclear Material}} \\
\cmidrule(lr){2-5} \cmidrule(lr){6-9}
\textbf{Method} & \textbf{TDR} & \textbf{TCR} & \textbf{TIDR} & \textbf{FPR} & \textbf{TDR} & \textbf{TCR} & \textbf{TIDR} & \textbf{FPR} \\
\midrule
CNN & \textbf{0.0219} & \textbf{0.0219} & \textbf{0.0219} & 0.0838 & 0.1345 & 0.1345 & 0.1201 & 0.0279 \\
MLP & 0.0088 & 0.0088 & 0.0088 & 0.0279 & 0.1094 & 0.1081 & 0.0918 & 0.0524 \\
ViT & 0.0000 & 0.0000 & 0.0000 & \textbf{0.0070} & 0.1018 & 0.1018 & 0.0968 & 0.0593 \\
Baseline NMF & 0.0000 & 0.0000 & 0.0000 & 0.0244 & \textbf{0.2879} & \textbf{0.2627} & \textbf{0.1565} & \textbf{0.0244} \\
\bottomrule
\end{tabular}
\end{subtable}

\end{table}

\end{document}